\documentclass[12pt]{iopart}

\usepackage{epsf}
\usepackage{citesort}

\newcommand{\illuseps}[2]{
\begin{figure}
\begin{center}
\epsfbox{#1.eps}
\caption{\label{#1}#2}
\end{center}
\end{figure}
}

\begin{document}

\title[Self-driven Monte Carlo simulations]{Self-driven lattice-model Monte Carlo simulations of alloy thermodynamic
properties and phase diagrams}
\author{A. van de Walle and M. Asta}
\address{Materials Science and Engineering Department,
Northwestern University, 2225 North Campus Drive, Evanston, IL 60208, USA}
\ead{avdw@alum.mit.edu}

\begin{abstract}
Monte Carlo (MC) simulations of lattice models are a widely used way to
compute thermodynamic properties of substitutional alloys. A limitation to
their more widespread use is the difficulty of driving a MC simulation in
order to obtain the desired quantities. To address this problem, we have
devised a variety of high-level algorithms that serve as an interface
between the user and a traditional MC code. The user specifies the goals
sought in a high-level form that our algorithms convert into elementary
tasks to be performed by a standard MC code. For instance, our algorithms
permit the determination of the free energy of an alloy phase over its
entire region of stability within a specified accuracy, without requiring
any user intervention during the calculations. Our algorithms also enable
the direct determination of composition-temperature phase boundaries
without requiring the calculation of the whole free energy
surface of the alloy system.
\end{abstract}

\submitto{Modelling Simul. Mater. Sci. Eng.}

\maketitle

\section{Introduction}

Monte Carlo (MC) simulations \cite{newman:mc,binder:mc,dunweg:mc,laradji:mc} of lattice models
are a widely used method to compute thermodynamic properties of
substitutional alloys. A lattice model MC simulation only requires, as an
input, a so-called \emph{cluster expansion} \cite
{sanchez:cexp,fontaine:clusapp,zunger:NATO}, which defines the energetics of
the system by specifying the energy associated with any atomic arrangement
on a given lattice. Cluster expansions are typically constructed from a fit
to the energies of a set of structures calculated, for instance, from
first-principles density-functional-theory calculations \cite
{jones:ldareview}, thus enabling the prediction of thermodynamic properties
even in the absence of experimental data.

In its simplest form, a cluster expansion specifies the energy difference
between nearest-neighbor chemical bonds joining identical and distinct
species in a binary alloy ($E_{AB}-\left( E_{AA}+E_{BB}\right) /2$) and
predicts compound formation energies from the number of each type of
nearest-neighbor bond. However, this formalism can be extended \cite
{sanchez:cexp}\ to allow for arbitrary interaction ranges and multibody
interactions so that the energetics of the alloy can be modeled with an
arbitrary accuracy by including a sufficient number of such interactions. It
has been demonstrated that the energetic contributions arising from
displacements of the atoms away from their ideal lattice sites \cite
{fontaine:soe,laks:recip,ducastelle:book,ceder:ising,fontaine:clusapp,zunger:NATO,wolverton:releci,asta:noble}
can be accounted for within this framework. Moreover, by allowing
temperature-dependent interactions, entropic contributions arising from
vibrational and electronic excitations can be included as well (see \cite
{avdw:vibrev} for a review and \cite
{ozolins:alsc,ozolins:cuau,wolverton:cual,wolverton:selec} for recent examples). Since
long-range interactions are often needed to properly describe the energetics
of an alloy system, MC simulations have become a preferred method to compute
thermodynamics properties from cluster expansions, because it would be
difficult to account for such long-range interactions within accurate
mean-field methods, such as the Cluster Variation Method (CVM) \cite
{kikuchi:cvm} while efficient algorithms to include even infinite-range interactions in MC
simulations have been devised \cite{lu:rsmc}. MC simulations are also conceptually simple to understand,
straightforward implement, and can be easily adapted to compute a wide variety
of properties.

Given these attractive features, it is not surprising that MC simulations of
lattice models with cluster expansions determined from first-principles
calculations have been employed in a variety of context for metallic,
semi-conductor and ceramic systems, including the computation of:
composition-temperature phase diagrams, thermodynamic properties of stable
and metastable phases, short-range order in solid solutions, thermodynamic
properties of planar defects (including surfaces or antiphase and interphase
boundaries), and the morphology of precipitate microstructures \cite
{ducastelle:book,fontaine:clusapp,zunger:NATO,zunger:scord,wolverton:srorev,ceder:oxides,asta:fppheq}%
.

With the steadily increasing availability of computational resources, MC
simulations of lattice models are likely to become even more prevalent in
computational material science modeling. Despite the technological advances
in computational hardware, the human time needed to select the appropriate
simulation parameters and to analyze the data has remained essentially
constant over the years. We are now in a situation where the main limitation
to the inclusion of atomistic MC simulations in standard thermodynamic
software toolkits is not the lack of computational resources, but rather the
difficulty of controlling a MC simulation in order to obtain the desired
quantities.

To address this problem, we have devised a variety of high-level algorithms
that serve as an interface between the user and a traditional MC code. The
user specifies the goals sought in a high-level form that our algorithm
converts into elementary tasks to be performed by a standard MC code. For
instance, our algorithms permit the determination of the free energy of a
phase over all of its region of stability within a specified accuracy,
without requiring any user intervention during the calculations. The
proposed algorithms have been implemented in an easy-to-use software package 
\cite{avdw:emc2code}.

In an effort to make this article self-contained, we first recall the basic
theoretical results underlying semi-grand-canonical MC simulations, which
are especially suited for determination of thermodynamic properties of
alloys. We then proceed to describe the algorithms that enable the automated
calculation of thermodynamic quantities from MC simulations. These
algorithms address the following three questions. (i) How to determine when
a MC simulation has run for a sufficiently long of time to provide the
desired quantities within a given target precision? (ii) How to detect phase
transitions? (iii) How to efficiently determine the composition-temperature
phase boundaries of a phase diagram? We then provide a series of examples of
applications that illustrate the features of our algorithms.

\section{Semi-Grand-Canonical Monte Carlo}

A very convenient way to obtain thermodynamic information regarding an alloy
system is to perform MC simulations which sample a semi-grand-canonical
ensemble, also known as the transmutation ensemble. In this ensemble, the
energy and concentration of an alloy with a fixed total number of atoms ($N$%
) are allowed to fluctuate while temperature and chemical potentials are
externally imposed. Although the results of this paper can be extended to
general multicomponent systems, we are considering the case of binary
alloys, for the benefit of notational simplicity. In an $A-B$ alloy, without
loss of generality, only the composition $x$ of element $A$ and the \emph{%
difference} in\ chemical potential between the two species $\mu =\mu
_{A}-\mu _{B}$ needs to be specified. For conciseness, we simply refer to
the quantity $\mu $ as the ``chemical potential'' and $x$ as the
concentration.

The natural thermodynamic potential $\phi $ (expressed per atom) associated
with the semi-grand-canonical ensemble can be defined in terms of the
partition function of the system as follows: 
\begin{equation}
\phi \left( \beta ,\mu \right) =-\frac{1}{\beta N}\ln \left( \sum_{i}\exp
\left( -\beta N\left( E_{i}-\mu x_{i}\right) \right) \right) 
\label{partfunc}
\end{equation}
where $E_{i}$ and $x_{i}$ are, respectively, the internal energy (per atom)
and the concentration of state $i$, while $\beta =1/\left( k_{B}T\right) $
is the reciprocal of the temperature $T$ and $k_{B}$ is Boltzmann's
constant. The thermodynamic potential $\phi $ is related to the Helmholtz
free energy $F$ through $\phi =F-\mu x$.

The potential $\phi $ can also be defined by the following total
differential:
\begin{equation}
d\left( \beta \phi \right) =\left( E-\mu x\right) \,d\beta -\beta x\,d\mu.
\label{totdiff}
\end{equation}
where $E$ and $x$ are the system's average internal energy (per atom) and
concentration. Equation (\ref{totdiff}) enables the calculation of the
thermodynamic function $\phi \left( \beta ,\mu \right) $ through
thermodynamic integration: 
\begin{equation}
\phi \left( \beta _{1},\mu _{1}\right) =\phi \left( \beta _{0},\mu
_{0}\right) +\frac{1}{\beta }\int_{\left( \beta _{0},\mu _{0}\right)
}^{\left( \beta _{1},\mu _{1}\right) }\left( E-\mu x,\,-\beta x\right) \cdot
d\left( \beta ,\mu \right) ,  \label{thint}
\end{equation}
where the integral is performed along a continuous path joining points $%
\left( \beta _{0},\mu _{0}\right) $ and $\left( \beta _{1},\mu _{1}\right) $
that does not encounter a phase transition. Semi-grand-canonical MC
simulations can be directly used to determine the values of $E$ and $x$ for
any given $\beta $ and $\mu $.

The potential at the initial point $\phi \left( \beta _{0},\mu _{0}\right) $
is usually chosen to be a convenient point where $\phi $ can be computed
analytically. One possible starting point is the high temperature limit \cite
{ducastelle:book},
\begin{equation}
\beta \phi \left( \beta ,\mu \right) \rightarrow -\ln \left( 2\right) \mbox{
as }\beta \rightarrow 0  \label{HTE}
\end{equation}
where the limit is taken while keeping $\mu $ finite (implying
that concentration $x$ converges to $1/2$ as $\beta \rightarrow 0$).
Another possible starting point is in the limit of low temperature at a
chemical potential stabilizing a given ground state $g$. In this limit, a
``single-spin flip'' low temperature expansion (LTE) \cite
{ducastelle:book,afk:lte,woodward:sgce} can be used to obtain: 
\begin{equation}
\phi \left( \beta ,\mu \right) =E_{g}-\mu x_{g}-\frac{1}{\beta N}%
\sum_{s}\exp \left( -\beta \left( \Delta \varepsilon _{s,g}-\mu \Delta \eta
_{s,g}\right) \right) ,  \label{LTE}
\end{equation}
where $E_{g}$ is the energy (per atom) of ground state $g$ with composition $%
x$, while $\Delta \varepsilon _{s,g}$ is the variation in the system's total
energy $\left( NE\right) $ associated with changing the identity of the atom
sitting at site $s$ in ground state $g$, and $\Delta \eta _{s,g}$ is the
variation in $\left( Nx\right) $ associated with the same change.
This infinite sum can be reduced to a finite number of terms by making
use of the translational periodicity of the ground state.

In order to avoid lengthy MC simulations, it is useful to possess a
criterion to find the highest possible temperature where the LTE is
accurate. One attractive possibility is to compute $\phi $ both from the LTE
and from a mean field approximation (MFA). As long as both approaches give
the same result within a user specified tolerance, we can be confident that
``multiple spin flip events'' are rare and that the LTE has the desired
level of precision.

Using the LTE for each ground state of the system as a starting point to
evaluate Equation (\ref{thint}), one can map out the whole potential surface 
$\phi ^{\alpha }\left( \beta ,\mu \right) $ associated with any given
ordered phase $\alpha $. Similarly, the high temperature limit can be used
as a starting point to obtain $\phi ^{\alpha }\left( \beta ,\mu \right) $
for the disordered phase. This process is illustrated in Figure \ref{exthint}%
.\ Once $\phi ^{\alpha }\left( \beta ,\mu \right) $ has been determined for
all phases, the boundary between two given phases $\alpha $ and $\gamma $
can be located by identifying the locus $\Lambda ^{\alpha \gamma }=\left\{
\left( \beta ,\mu \right) :\phi ^{\alpha }\left( \beta ,\mu \right) =\phi
^{\gamma }\left( \beta ,\mu \right) \right\} $ where the potential surfaces
intersect. At such intersections $\left( \beta ,\mu \right) \in \Lambda
^{\alpha \gamma }$, the concentrations $x^{\alpha }$ and $x^{\gamma }$ of
the two phases in equilibrium are simply given by the slopes of the
potential surface at the point of intersection: 
\begin{eqnarray}
x^{\alpha } &=&-\frac{\partial \phi ^{\alpha }\left( \beta ,\mu \right) }{%
\partial \mu } \\
x^{\gamma } &=&-\frac{\partial \phi ^{\gamma }\left( \beta ,\mu \right) }{%
\partial \mu }.
\end{eqnarray}

\illuseps{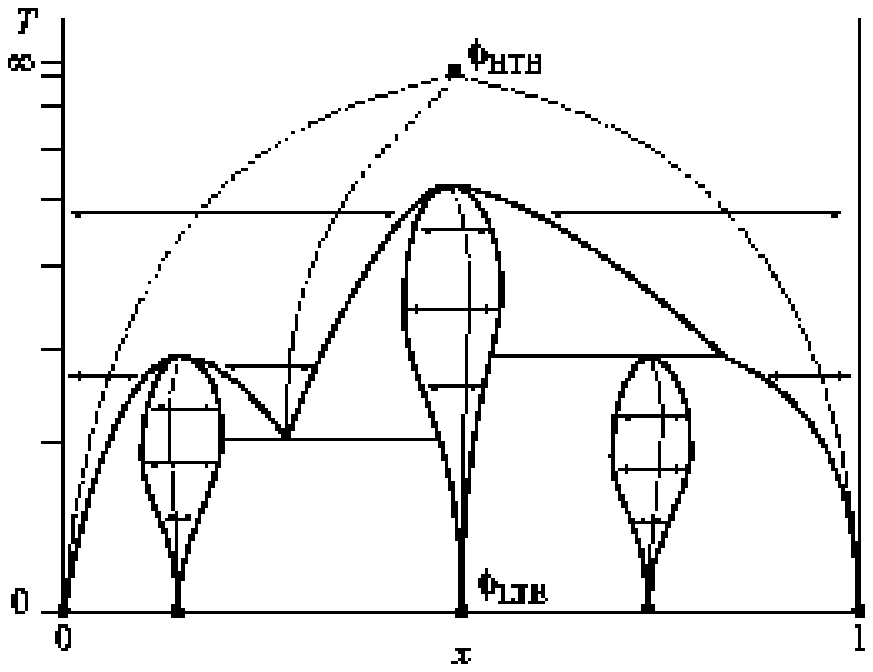}{Example of paths of integration to determine
the potential $\phi \left( \beta ,\mu \right) $ of every phase. The starting
points are given by either the Low Temperature Expansion (LTE) or the High
Temperature Expansion (HTE).}

The calculations of thermodynamic quantities based on semi-grand-canonical
MC simulations offer several advantages over their canonical counterparts.

\begin{enumerate}
\item  For a given value of the control variables $\left( \beta ,\mu \right) 
$, the thermodynamic equilibrium of the system is never a phase-separated
mixture, implying that the
calculated quantities always reflect the properties of pure phases, free of
interfacial contributions that would otherwise not be negligible due to the
finite size of the simulation cell. (Note that the MC simulation cell must be
commensurate with the periodicity of the equilibrium structure for this
property to hold.)

\item  The potential $\phi $ can be very directly determined through a
simple thermodynamic integration procedure (Equation (\ref{thint})) where
each required quantity takes the form of a thermodynamic average of
computationally inexpensive quantities ($E$, which must already be computed
in order to perform the simulation, and $x$, which is trivial to obtain).

\item  Phase boundaries can be located by looking for intersections between
curves, a criterion which is somewhat simpler to implement than the common
tangent construction (although both methods are formally equivalent).
\end{enumerate}

\section{Algorithms}

While the methods described in the previous section in principle enable the
determination of thermodynamic quantities from MC simulations, a number of
practical issues need to be addressed before\ this process can be fully
automated.

A typical MC code requires a substantial amount of user intervention and
relies on the user's physical intuition to identify when a simulation is
successful and reliable. The type of user interventions needed include:

\begin{enumerate}
\item  Providing a variety of input parameters controlling the computational
accuracy that need to be determined by trial and error.

\item  Requiring the user to monitor the simulation periodically to ensure
that the system has not undergone an unexpected phase transformation.

\item  Performing a substantial amount of post-processing in order to obtain
a phase diagram.
\end{enumerate}

In this section we introduce algorithms to perform these tasks
automatically and provide a formal justification of their applicability.

\subsection{Reaching a target precision\label{autoprec}}

Three parameters control the precision of a thermodynamic quantity computed
through MC simulations: the simulation cell size, the equilibration time and
the averaging time. While the analysis of the effect of simulation cell size
on the precision is beyond the scope of the present paper, this section
provides a sound basis for the selection of the two remaining parameters. The
equilibration time is the time the system takes to reach thermodynamic
equilibrium from a given initial configuration. Once equilibrium is reached,
the instantaneous value of a given quantity (e.g. the energy) then needs to
be averaged over a certain number of MC steps in order for its average to be
sufficiently accurate. This defines the averaging time, which is the first
parameter we will focus on.

\subsubsection{Averaging time}

While the approach presented in this section relies on standard statistical
results, it is nevertheless included for the purpose of introducing our
notation and clarifying the assumptions made. Consider a microscopic
quantity taking the value $Q_{t}$ at step $t$ and whose expectation value is
equal to the desired thermodynamic quantity $Q$. If $L$ observations of $%
Q_{t}$ are available, a natural estimator of $Q$ is the average: 
\begin{equation}
\overline{Q}_{\left[ 1,L\right] }=\frac{1}{L}\sum_{t=1}^{L}Q_{t}.
\end{equation}
The precision of this statistical quantity can be quantified\ by calculating
its variance. In general, the variance of $\overline{Q}_{\left[ 1,L\right] }$
is given by: 
\begin{equation}
Var\left( \overline{Q}_{\left[ 1,L\right] }\right) =\frac{1}{L^{2}}%
\sum_{t=1}^{L}\sum_{s=1}^{L}V_{\left| t-s\right| }  \label{genvar}
\end{equation}
where $V_{\left| t-s\right| }$ is the covariance between $Q_{t}$ and $Q_{s}$%
. The fact that the sequence $Q_{t}$ is stationary when the system has
reached thermodynamic equilibrium enables a simple estimator $\hat{V}_{l}$
of the covariance structure $V_{l}$ of the sequence through the equation 
\begin{equation}
\hat{V}_{l}=\frac{1}{L-l}\sum_{t=1}^{L-l}Q_{t}Q_{t+l}-\overline{Q}_{\left[
1,L\right] }^{2}.  \label{vhat}
\end{equation}
While Equation \ref{genvar} is very general, it is also very computationally
intensive, requiring on the order of $L^{2}$ operations to estimate the
variance of $\overline{Q}_{\left[ 1,L\right] }$. Following most of the
literature on MC simulation, we consider the more computationally tractable
special case obtained when the covariance structure $V_{l}$ is assumed to be
of the form 
\begin{equation}
V_{l}=V\rho ^{-\left| l\right| }.
\end{equation}
This approximation can be motivated as follows. A Markov chain can be
represented by a linear superposition of components with exponentially
decaying autocorrelation function. Each component is characterized by a
certain correlation length and the component with the longest correlation
length will typically give by far the largest contribution to the variance.
Considering only the most slowly decaying component should thus provide a
good approximation to the true variance of the quantity of interest.

Under the assumption that the averaging time is much larger than the longest
correlation time,\footnote{%
This assumption should not be a concern because (i) this assumption tends to be
violated only for averaging times such that the desired target precision on $%
\overline{Q}$ has not yet been reached, and (ii) our expression
overestimates the true variance when this assumption is violated.}
``boundary'' effects can be neglected and Equation (\ref{genvar}) then
reduces to 
\begin{equation}
Var\left( \overline{Q}_{\left[ 1,L\right] }\right) =\frac{1}{L^{2}}%
\sum_{t=1}^{L}\sum_{l=-\infty }^{\infty }V\rho ^{\left| l\right| }.
\end{equation}
Evaluating the geometric sum then yields: 
\begin{equation}
Var\left( \overline{Q}_{\left[ 1,L\right] }\right) =\frac{V}{L}\left( \frac{%
1+\rho }{1-\rho }\right) .  \label{varar}
\end{equation}
Note that, since $\rho =\exp \left( -1/\tau \right) $ where $\tau $ is the
characteristic relaxation time, Equation (\ref{varar}) can be written, in
the limit of large $\tau $, in the more familiar form $Var\left( \overline{Q}%
_{\left[ 1,L\right] }\right) =2\tau V/L$.

While $V$ can be directly estimated from Equation (\ref{vhat}), setting $l=0$%
, the parameter $\rho $ can be found in a variety of ways. We found the
following algorithm to be both fast and accurate: Find the time interval 
$l^{\ast }$ such that $\hat{V}_{l^{\ast }}/\hat{V}_{0}=\frac{1}{2}$ and then
set $\rho = 2^{-1/l^{\ast}}$. This approach offers the
important advantage that it uses, as an input,
only the correlation between relatively distant values of $Q_{t}$. This
ensures that $\rho $ will be determined by the most slowly decaying
components of the Markov process, which are the ones that the functional
form $V\rho ^{\left| l\right| }$ was meant to model. It is also a very fast
algorithm, since a simple bracketing algorithm allows the determination of $%
\rho $ in about $L\log L$ operations.
In the event that there are more than one $l^{\ast }$ satisfying the
equality, we chose the smallest one. The rationale behind this choice
is that there is a small but nonzero
chance that the estimated covariances are such that
$\hat{V}_{l}/\hat{V}_{0}=1/2$ although the true covariances are \emph{not} such
that $V_{l}/V=1/2$. Choosing the smallest $l^{\ast }$ makes the
procedure less sensitive to these spurious events since the
probability of occurence of one such event increases with the range of
values of $l$ considered.

In short, the averaging time $L$ needed to reach a given precision can be
found by computing the variance of $\overline{Q}_{\left[ 1,L\right] }$
through Equation (\ref{varar}) periodically and stopping the simulation when
the variance goes below a user specified threshold: 
\begin{equation}
Var\left( \overline{Q}_{\left[ 1,L\right] }\right) \leq \left( \frac{p}{%
z_{\alpha }}\right) ^{2},  \label{varcrit}
\end{equation}
where $p$ is the target precision on the value of $Q$ while 
\begin{equation}
z_{\alpha }=\sqrt{2}\,\mbox{erf}^{-1}\left( 1-\alpha \right)   \label{zalpha}
\end{equation}
is the critical value associated with the desired level of confidence $%
\alpha $. For instance, if the error on $Q$ is required to be less than $%
10^{-3}$ with a probability of $99\%$, then $p=10^{-3}$, $\alpha =0.01$ and $%
z_{\alpha }=2.576$.

\subsubsection{Equilibration time}

When devising a criterion to determine whether a simulation has reached
thermodynamic equilibrium, approaches based on identifying correlation
lengths can be unreliable because the system may be very far from
equilibrium during equilibration. For this reason, we employ a relatively
simple, and yet very robust approach that is based on the main defining
characteristic of thermodynamic equilibrium. If the average of the quantity $%
Q_{t}$ between steps $t_{1}$ and $t_{2}$ is not statistically significantly
different from the average of $Q_{t}$ between steps $t_{2}$ and $t_{3}$,
then the hypothesis that equilibrium has been reached at step $t_{1}$ and
beyond cannot be rejected. Of course, the power of this test increases as $%
t_{2}-t_{1}$ and $t_{3}-t_{2}$ increase and this definition of equilibrium
thus depends on the target precision $p$ we seek to reach. This is to be
expected since, strictly speaking, thermodynamic equilibrium is never
reached in a simulation of a finite length. Any definition of equilibrium
must include an arbitrary cutoff and it is natural to choose this cutoff to
be a function of the precision we require.

Both the equilibration and averaging time needed to reach a precision $p$
can thus be determined through an algorithm that can be outlined as follows
(refer to Figure \ref{eqcrit}). Let $L$ denote the number of samples of $%
Q_{t}$ collected so far. Consider a range of values of $t_{1}\in \left[ 1,L%
\right] $ and for each, check whether 
\begin{equation}
\Delta \overline{Q}\equiv \left| \overline{Q}_{\left[ t_{1},\left(
t_{1}+L\right) /2\right] }-\overline{Q}_{\left[ \left( t_{1}+L\right) /2,L%
\right] }\right| \leq p,
\end{equation}
where $Q_{\left[ t_{1},t_{2}\right] }$ denotes the average of $Q_{t}$ for $%
t\in \left[ t_{1},t_{2}\right] $.\ Then, verify that $Var\left( \overline{Q}%
_{\left[ t_{1},L\right] }\right) $ satisfies Equation (\ref{varcrit}). If
both conditions are satisfied, then $\overline{Q}_{\left[ t_{1},L\right] }$
provides an estimate of $Q$ with a precision $p$ and with a confidence level 
$\alpha $.

\illuseps{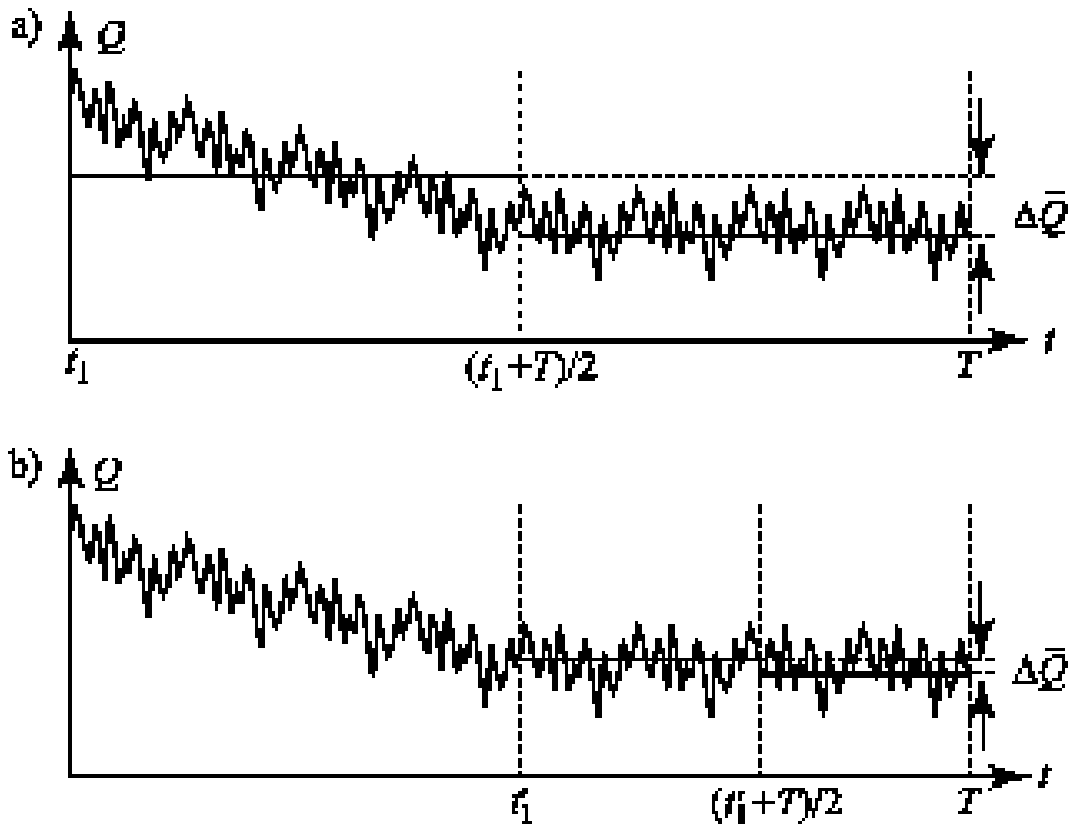}{Criterion for testing whether thermodynamic
equilibrium has been reached in a MC simulation. In part a), the two
blocks of data have significantly different means, implying that the
hypothesis that equilibrium has been reached at $t_{1}$ is
rejected. In part b), the two blocks of data do have similar means,
indicating that the hypothesis that equilibrium has been reached at
$t_{1}^{\prime }$ cannot be rejected.}

In its simplest form, the implementation of this algorithm involves storing 
\emph{all} samples of $Q_{t}$ collected so far and requires a loop over
values of $t_{1}$ which make the complexity of the algorithm $O\left(
L^{2}\right) $. Various techniques help reduce the resources required.
First, sampling $Q_{t}$ every MC pass,\footnote{%
A MC pass consists of a number of attempted spin flips equal to the number
of lattice sites in the simulation cell.} instead of at every spin flip,
results in essentially no loss in precision since the correlation length of
the simulation typically extends much beyond a MC pass. Second, one can
limit the search for $t_{1}$ to values that are integer multiple of some
block size $L_{b}$ that is allowed to increase linearly with the total
number of MC passes $L$. In this fashion, the number of operations needed to
determine $t_{1}$ does not increase as the simulation progresses. The value
of $t_{1}$ thus found by a search constrained to block boundaries differs at
most by $L_{b}$ from the $t_{1}$ which would have been found without block
constraints. In the worst case, the simulation may have to be run for an
additional $L_{b}$ passes and it follows that the length of the simulation
will be at most $\left( 1+L_{b}/L\right) $ larger than the optimal
computational time that would have been needed without block constraints. A
computationally convenient way to implement this scheme is to multiply the
block size $L_{b}$ by two whenever the total number of passes $L$ doubles.
Thanks to this approach, the entire history of values of $Q_{t}$ does not
need to be stored, only block averages of $\overline{Q}$, $V$ and $\rho $
are needed, as illustrated in Figure \ref{boxalgo}. When the block size
doubles, the values of $\overline{Q}$, $V$ and $\rho $ for two consecutive
blocks can be combined to obtain the corresponding value for the new block
having twice the length.\footnote{%
The only drawback is that the resulting estimate of $\rho $ does not make
use of all the available data points, because the product $Q_{t}Q_{s}$ for $%
t $ and $s$ belonging to different blocks cannot be determined from
within-block averages. Although it is inefficient, our estimate of $\rho $
is nevertheless unbiased and the loss of efficiency is negligible if the
block size is large relative to the correlation time.}

\illuseps{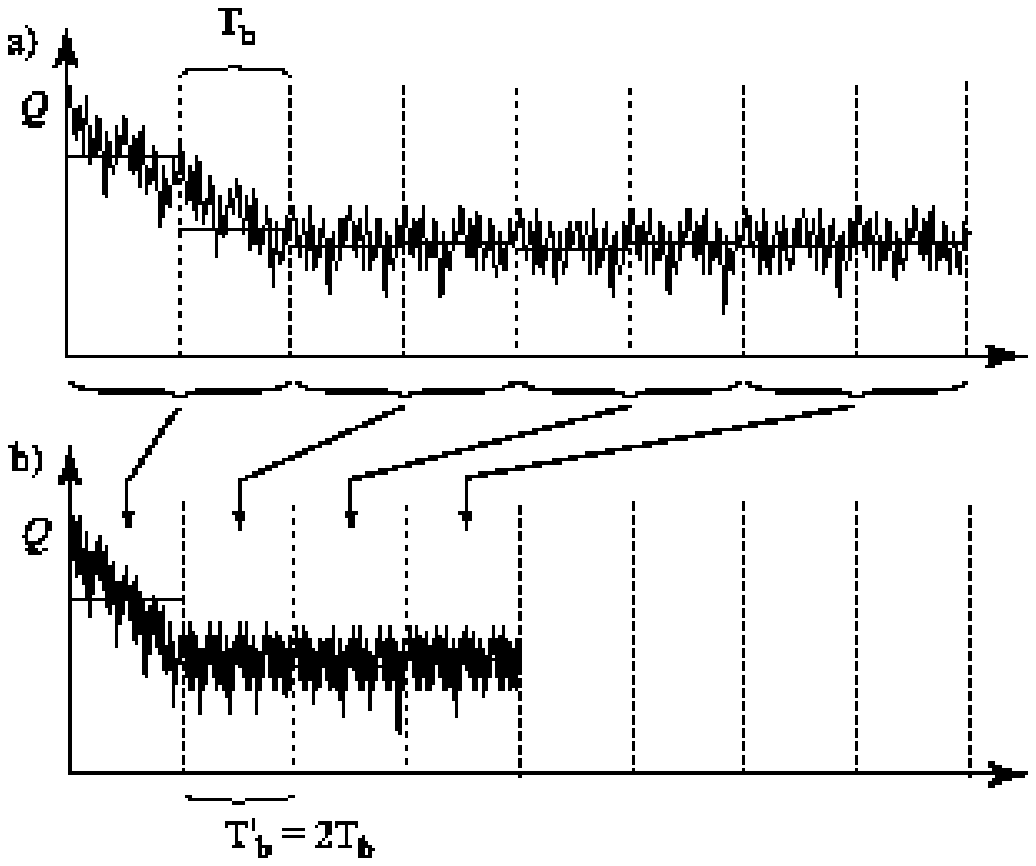}{Block-wise storage scheme. a) Only the mean
variance and correlations within blocks of width $L_{b}$ needs to be stored
(the mean is represented by a thick horizontal line). b) As the simulation
runs, blocks are periodically merged so that storage requirements do not
grow.}

\subsection{Detecting phase transitions}

\label{phtcv}Phase transitions can be identified by locating singularities
in some function $Q\left( C\right) $, where $Q$ is some thermodynamic
quantity (such as energy, concentration or an order parameter) and $C$ is
some control variable (such as temperature or chemical potential). In this
section, we propose a simple algorithm to perform the detection of such
singularities while accounting for the fact that quantities obtained from MC
are intrinsically noisy. Our algorithm is applicable to the detection of
both first-order and second-order transitions, although the power of the test
is clearly higher for first-order transitions.

An analytic function has the property that the knowledge of its shape in
some finite interval enables the prediction of its value outside of that
interval. The failure of this extrapolation procedure in the presence of a
singularity permits its detection. Formally, an analytic function can be
extrapolated using a Taylor series. We approximate this operation by fitting
the output of the MC simulation by a polynomial.

Consider $n+1$ estimates $\overline{Q}_{1},\ldots ,\overline{Q}_{n+1}$\ of
the quantity $Q\left( C\right) $ obtained for a sequence of values of the
control variable $C_{1},\ldots ,C_{n+1}$. The general principle behind our
singularity detection procedure is to fit a polynomial through the first $n$
points and if the observed value $\overline{Q}_{n+1}$ is far from the value
predicted from the polynomial fitted to the previous data, a singularity
must be present between point $n$ and $n+1$ (see Figure \ref{discrit}).\ To
make this algorithm practical, two questions need to be answered. First, how
do we choose the order of the polynomial to be fitted? Second, how do we
select the maximum allowable prediction error that can be tolerated without
claiming the presence of a singularity?

\illuseps{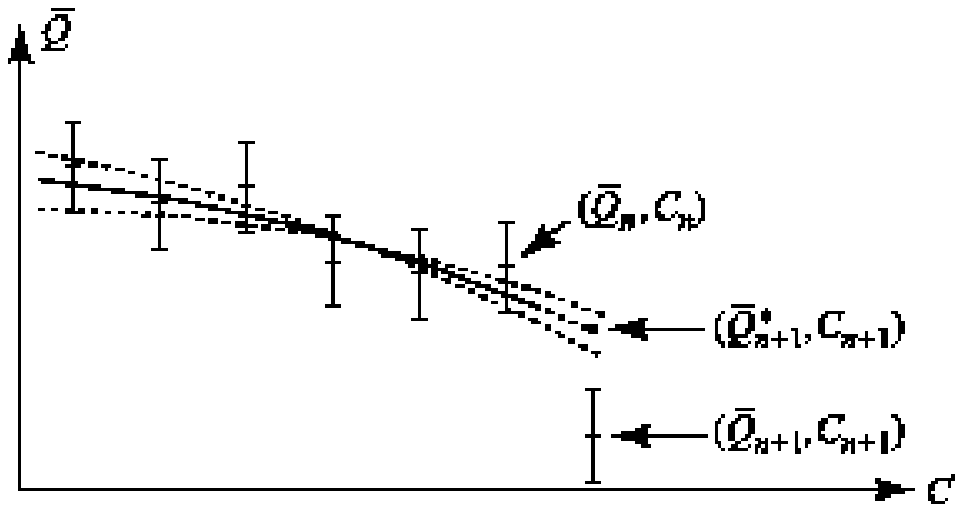}{Criterion for detecting a phase transition.
If the sequence $\left( \overline{Q}_{n},C_{n}\right) $ predicts a value of $%
\left( \overline{Q}_{n+1}^{\ast },C_{n+1}\right) $ but the actual value, $%
\left( \overline{Q}_{n+1},C_{n+1}\right) $ is statistically significantly
different, then a phase transition must have occurred. Both the noise in the
prediction (represented by dashed lines) and the noise in $\overline{Q}%
_{n+1} $ (represented by an error bar) must be taken into account to
determine statistical significance.}

To answer the first question, we make use of the well-known cross-validation
criterion \cite{stone:cv,li:cv}: Choose the order $k$ of the polynomial that
minimizes the quantity: 
\begin{equation}
CV=\frac{1}{n}\sum_{i=1}^{n}\left( \overline{Q}_{i}-Q_{i}^{\ast }\right) ^{2}
\end{equation}
where $Q_{i}^{\ast }$ is the value of $\overline{Q}_{i}$ predicted from a
polynomial least-squares fit to the remaining $n-1$ points: $\overline{Q}%
_{1},\ldots ,\overline{Q}_{i-1},\overline{Q}_{i+1},\ldots ,\overline{Q}_{n}$%
. Intuitively, this criterion tests the predictive power of the polynomial
using the points $\overline{Q}_{1},\ldots ,\overline{Q}_{n}$ before we
attempt to use it for the prediction of $\overline{Q}_{n+1}$. Once the order 
$k$ of the polynomial has been chosen, the standard theory of least-squares 
\cite{goldberger:ols}\ provides the distribution of the predicted value $%
Q_{n+1}^{\ast }$, which can be used to determine whether the prediction
error is statistically significantly different from zero. For $i=1,\ldots ,n$
and $j=1,\ldots ,k+1$, let 
\begin{eqnarray}
X_{ij} &=&\left( C_{i}\right) ^{j-1} \\
Y_{i} &=&\overline{Q}_{i},
\end{eqnarray}
so that the equation defining the least-squares problem can be written in
matrix form as $Y=Xa+\varepsilon $, where $\varepsilon $ is an $n\times 1$
vector of error terms. The vector $a$ of the coefficients of the polynomial
can be estimated by 
\begin{equation}
\hat{a}=\left( X^{T}X\right) ^{-1}X^{T}Y  \label{bols}
\end{equation}
while the covariance matrix of $\hat{a}$ is given by 
\begin{equation}
V=\sigma ^{2}\left( X^{T}X\right) ^{-1}  \label{viid}
\end{equation}
where $\sigma ^{2}$ is the variance of the residuals $\varepsilon _{i}$ of
the fit. The quantity $\sigma ^{2}$ can be estimated in two asymptotically
equivalent ways. Equation (\ref{varar}) provides an estimate of the variance
of any of the $\overline{Q}_{i}$, which can be averaged over $i$ to yield a
single, more accurate estimate: 
\begin{equation}
\hat{\sigma}^{2}=\frac{1}{n}\sum_{i=1}^{n}Var\left( \overline{Q}_{i}\right) .
\label{avgvarq}
\end{equation}
Alternatively, the residuals of the least-squares fit can be used: 
\begin{equation}
\hat{\sigma}^{2}=\left\| Y-Xa\right\| ^{2}/\left( n-k-1\right) .
\label{sigmahat}
\end{equation}
While Equation (\ref{avgvarq}) is more accurate (especially if $n$ is
small), because it makes use of additional information, Equation (\ref
{sigmahat}) is simpler to implement as it minimizes the amount of
information that needs to be transferred between the Monte Carlo simulation
code and the phase transition detection routine.

The predicted value of $Q$ for point $n+1$ is then given by 
\begin{equation}
Q_{n+1}^{\ast }=x^{T}\hat{a}
\end{equation}
where $x_{j}=\left( C_{n+1}\right) ^{j-1}$ for $j=1,\ldots ,k+1$, while the
variance of this prediction is given by: 
\begin{equation}
v=x^{T}Vx.
\end{equation}
If there is no singularity, the variance of the difference between $%
\overline{Q}_{n+1}$ and $Q_{n+1}^{\ast }$ is $v+\sigma ^{2}\ $because (i)
the variance of $Q_{n+1}^{\ast }$ is $v$ while the variance of $\overline{Q}%
_{n+1}$ is $\sigma ^{2}$, (ii) $\overline{Q}_{n+1}$ and $Q_{n+1}^{\ast }$
are independent. An asymptotically valid statistical test thus consists in
rejecting the hypothesis that there is no phase transition if the prediction
error $\left| \overline{Q}_{n+1}-Q_{n+1}^{\ast }\right| $ is such that%
\footnote{%
Note that $Q_{n+1}^{\ast }$ must be an unbiased predictor of $Q\left(
C_{n+1}\right) $ for this test to be valid, which is not the case for finite 
$n$ due to the finite order of the approximating polynomial. Fortunately,
the bias (squared) goes to zero as $n\rightarrow \infty $ and eventually
becomes negligible relative to $\sigma ^{2}$, ensuring the asymptotic
validity of the procedure. Note that the variance $v$ of the prediction is
also negligible asymptotically relative to $\sigma ^{2}$. We nevertheless
include the variance term $v$ in Equation (\ref{critsing}) in order to
improve the accuracy of the test for finite $n$, since we do have an
estimator of the variance (unlike the bias).} 
\begin{equation}
\left| \overline{Q}_{n+1}-Q_{n+1}^{\ast }\right| \geq z_{\alpha }\sqrt{%
v+\sigma ^{2}}.  \label{critsing}
\end{equation}
For $z_{\alpha }$ defined as in Equation (\ref{zalpha}), the probability
that a phase transition is incorrectly identified is less than $\alpha $.
The relevant quantities are represented in Figure \ref{discrit}.

The Equations presented above rely on the assumption that the deviations
away from the true thermodynamic quantity, $\left( \overline{Q}_{i}-Q\left(
C_{i}\right) \right) $ for $i=1,\ldots ,n+1$, are statistically independent.
This is an appropriate assumption if the system is given sufficient time to
equilibrate every time the value of the control variable $C$ changes. Our
method also assumes that the deviations $\left( \overline{Q}_{i}-Q\left(
C_{i}\right) \right) $ all have the same variance. Both of these
requirements are automatically satisfied if the points $\overline{Q}%
_{1},\ldots ,\overline{Q}_{n+1}$ are generated using the algorithm presented
in Section \ref{autoprec}.

For the sole purpose of locating phase transitions, one particular type of
thermodynamic function enables an especially sensitive test for phase
transitions: order parameters. Unfortunately, order parameters that are
invariant under any possible symmetry operation of the lattice are often
expensive to compute. However, if one always traverses phase transitions
from an ordered phase toward a disordered phase (or another ordered phase),
the task is dramatically simplified. When a MC cell is initialized with one
specific ground state (i.e. an ordered phase) and before a phase transition
is reached, there is a vanishing probability that the system will fluctuate
enough to reach a state that is different but symmetrically equivalent to
the original state. Thus, an order parameter can be computed by simply
keeping track of the number of sites that host an atomic specie that is
different from the specie found at the same site at initialization. There is
no need to check for every possible symmetry-related ordered configuration.

The algorithms just presented are especially useful when one desires to
construct the potential surface $\phi \left( \beta ,\mu \right) $ of a phase
(using thermodynamic integration over the paths depicted in Figure \ref
{exthint}) without requiring the user to manually determine when to stop the
thermodynamic integration along a given direction because a phase transition
has occurred. The automated construction of such free energy surfaces provides a
very natural pathway through which an atomistic MC simulation can be used to provide
input to CALPHAD-type calculations based on phenomenological free energy
models.

\subsection{Phase boundary tracing}

\label{phdtr}When one is solely interested in determining a system's phase
diagram, it would be computationally advantageous to be able to follow the
boundary of a phase without first calculating the potential $\phi $ over the
whole region where this phase is stable. In this section, we describe how,
in the case of a first-order transition, the entire boundary of a two-phase
equilibrium\ can be determined, starting from a single point where a
transition is known to occur. We first focus on a simple method that
neglects the unavoidable presence of statistical noise in MC simulations
before describing how the presence of this noise can be handled.

Consider a value of the reciprocal temperature $\beta $ and of the chemical
potential $\mu $ that is known to stabilize a phase-separated mixture
between phases $\alpha $ and $\gamma $. The knowledge of such a pair $\left(
\beta ,\mu \right) $ could be provided, for instance, by the phase
transition detection algorithm presented in the previous section, assuming
that the hysteresis loop of the phase transition is sufficiently narrow (which
tends to be case at high temperature). Alternatively, if $T=0$, the chemical
potential that simultaneously stabilizes two ground states $\alpha $ and $%
\gamma $ can be found by a simple common tangent construction from the
knowledge of the energy and composition of each ground state. At such a
two-phase equilibrium point, the potentials $\phi $ of each phase are equal.
Our task is to find, as reciprocal temperature $\beta $ changes by an amount 
$d\beta $, the changes in chemical potential needed to preserve the equality
between each phase's thermodynamic potential. By iterating this process, one
can define a whole path $\mu \left( \beta \right) $ that traces out the
value of the chemical potential at the phase transition as temperature
changes, and thus find the concentrations $x_{\alpha }$ and $x_{\gamma }$ of
each phase in equilibrium as a function of temperature. The differential
equation whose solution provides the desired path $\mu \left( \beta \right) $
can be found by taking the total differential of the equation $\beta \phi
^{\alpha }=\beta \phi ^{\gamma }$ following from the identity between the
potentials $\phi $ of each phase: 
\begin{eqnarray}
\left. \frac{\partial \left( \beta \phi ^{\alpha }\right) }{\partial \beta }%
\right| _{\mu }d\beta +\left. \frac{\partial \left( \beta \phi ^{\alpha
}\right) }{\partial \mu }\right| _{\beta }d\mu =\left. \frac{\partial \left(
\beta \phi ^{\gamma }\right) }{\partial \beta }\right| _{\mu }d\beta +\left. 
\frac{\partial \left( \beta \phi ^{\gamma }\right) }{\partial \mu }\right|
_{\beta }d\mu  \\
\Rightarrow \left( E^{\alpha }-\mu x^{\alpha }\right) d\beta -\beta
x^{\alpha }d\mu =\left( E^{\gamma }-\mu x^{\gamma }\right) d\beta -\beta
x^{\gamma }d\mu  \\
\Rightarrow \frac{d\mu }{d\beta }=\frac{\left( E^{\gamma }-E^{\alpha
}\right) }{\beta \left( x^{\gamma }-x^{\alpha }\right) }-\frac{\mu }{\beta },
\label{dmudT}
\end{eqnarray}
where $E^{\alpha }$ and $x^{\alpha }$ are the energy and concentration of
phase $\alpha $ at reciprocal temperature $\beta $ and chemical potential $%
\mu $ (and similarly for phase $\gamma $). All the quantities needed can be
directly obtained from MC simulations. Thanks to the fact that both phases
are metastable for values of the chemical potential in the neighborhood of
the true phase transition, it is possible to obtain $E$ and $x$ for both
phases at the same chemical potential. A very efficient way to implement
this algorithm is to keep two simulations in memory at the same time --- one
for the phase $\alpha $ and one for the phase $\gamma $. In this fashion,
the final configuration of the simulation of phase $\alpha $ at reciprocal
temperature $\beta $ can be used as an initial configuration for the
simulation of phase $\alpha $ at temperature $\beta +d\beta $ (and similarly
for phase $\gamma $). The long equilibration time associated with the
crossing of a phase transition is thus avoided. Interestingly, this
algorithm can be directly used even when the two phases in equilibrium are
based on a completely different parent lattice.

We now address the issue that the quantities obtained from MC are
contaminated by noise. The presence of noise causes the solution to Equation
(\ref{dmudT}) calculated from MC to deviate from the true path $\mu \left(
\beta \right) $. This eventuality brings up two questions. First, is the
solution to Equation (\ref{dmudT}) stable or unstable with respect to the
presence of small perturbations? Second, what can one do if the noise
becomes so large that the calculated chemical potential leaves the region
where both phases are metastable?

To answer the first question, consider a perturbation $\varepsilon \left(
\beta \right) $ of the chemical potential away from its value $\mu \left(
\beta \right)$ at the phase transition at reciprocal temperature $\beta $.
We now calculate how this error is further propagated when solving Equation 
\ref{dmudT}. To the first order, the difference between the potential $\phi $
of the two phases is $\beta \left( x^{\gamma }\left( \beta \right)
-x^{\alpha }\left( \beta \right) \right) \varepsilon \left( \beta \right) $.
As temperature changes, Equation (\ref{dmudT}) updates $\mu $ so as to
preserve this difference. We thus have the equality: 
\begin{equation}
\left( x^{\gamma }\left( \beta \right) -x^{\alpha }\left( \beta \right)
\right) \varepsilon \left( \beta \right) =\left( x^{\gamma }\left( \beta
+d\beta \right) -x^{\alpha }\left( \beta +d\beta \right) \right) \varepsilon
\left( \beta +d\beta \right) .
\end{equation}
Taking the limit of infinitesimal $d\beta $ and rearranging yields: 
\begin{equation}
\frac{d\varepsilon }{d\beta }=-\varepsilon \frac{d}{d\beta }\ln \left|
x^{\gamma }-x^{\alpha }\right|
\end{equation}
and it follows that the small perturbation $\varepsilon $ is unstable as $%
\beta $ increases when the difference in concentration between the two
phases in equilibrium $\left| x^{\gamma }-x^{\alpha }\right| $ decreases
with increasing $\beta $. The practical consequence of this observation is
that integrating Equation (\ref{dmudT}) from high temperature to low
temperature yields a method more robust to statistical noise, because $%
\left| x^{\gamma }-x^{\alpha }\right| $ usually increases with decreasing
temperature $T$.

We now address the second robustness issue, namely, what can we do if the
algorithm wanders outside of the region of metastability of either phase? It
is straightforward to detect when this happens by simply checking whether $%
\left| x^{\gamma }-x^{\alpha }\right| $ is not statistically significantly
different from zero. One then needs to determine which of the two MC
simulations (for phase $\alpha $ or $\gamma $) underwent a phase transition.
This can be accomplished by keeping track of the concentration of each phase
($x_{1}^{\alpha },\ldots ,x_{n}^{\alpha }$ and $x_{1}^{\gamma },\ldots
,x_{n}^{\gamma }$) at temperatures previously visited ($\beta _{1},\ldots
,\beta _{n}$) and using them the verify whether the new concentration $%
x_{n+1}^{\alpha }\approx x_{n+1}^{\gamma }$ at $\beta _{n+1}$ is best
predicted by the sequence $x_{i}^{\alpha }$ or $x_{i}^{\gamma }$, using the
extrapolation described in the previous section (see Figure \ref{phtcrit}).
The sequence that gives the worst prediction will point to the MC simulation
that underwent a phase transition.

\illuseps{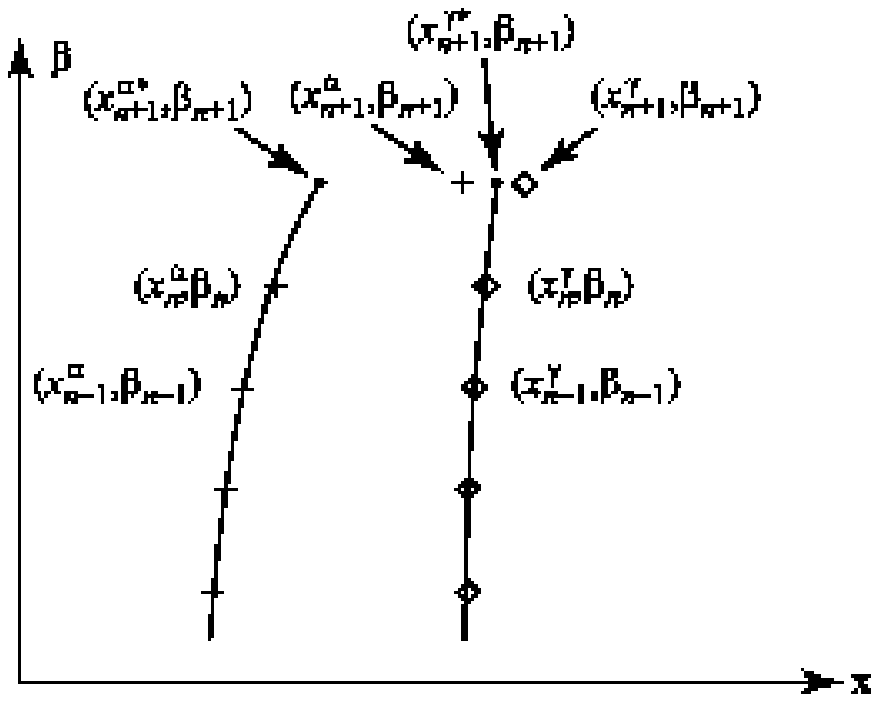}{Criterion for detecting when the phase
boundary tracing algorithm has exited the region of simultaneous
metastability of the two phases in presence. The point $\left(
x_{n+1}^{\alpha },\beta _{n+1}\right) $ is better predicted by $\left(
x_{n+1}^{\gamma \ast },\beta _{n}\right) $ than by $\left( x_{n+1}^{\alpha
\ast },\beta _{n}\right) $. Hence, the MC simulation of phase $\alpha $ must
have undergone a transition to phase $\gamma $.}

Without loss of generality consider the case when $x^{\alpha }<x^{\gamma }$
and when the simulation of phase\ $\alpha $ transformed to phase $\gamma $.
We can then ``recenter'' the chemical potential to a good estimate of its
true value at the phase transition as follows (refer to Figure \ref{recenter}%
).

\illuseps{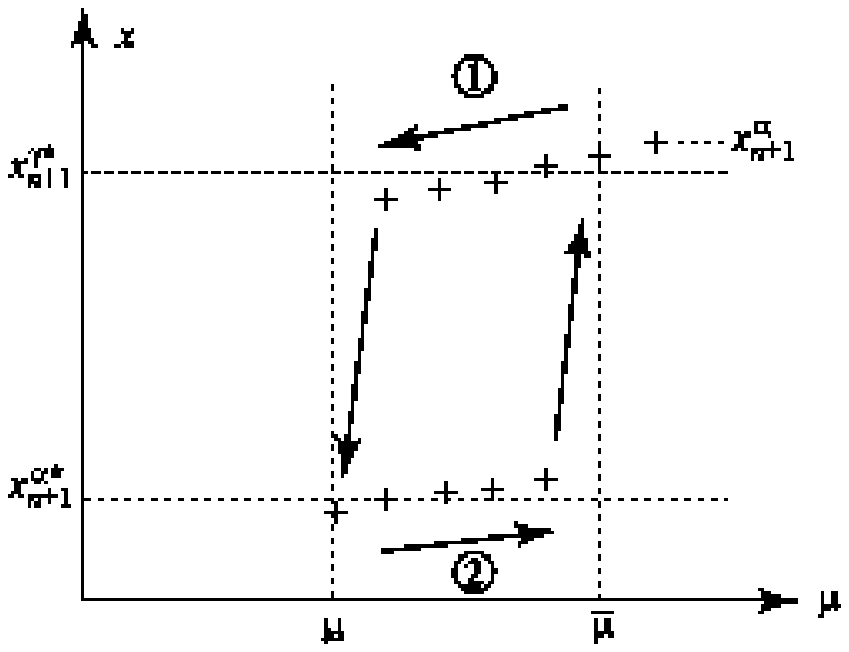}{Algorithm to locate the center of the
region of simultaneous metastability of the two phases in presence. In step
1, the chemical potential $\mu $ in the simulation of phase $\alpha $ is
gradually decreased until concentration is once again best predicted by $%
x_{i}^{\alpha \ast }$. In step 2, the chemical potential is increased until
the simulation transform back to phase $\gamma $. The two extreme values of
the chemical potential $\underline{\mu }$ and $\overline{\mu }$ thus defined
bracket the true value of $\mu $ at the phase transition.}

\begin{enumerate}
\item  Gradually decrease the chemical potential of the MC simulation of
phase $\alpha $ until its concentration can again be predicted by the
sequence $x_{i}^{\alpha }$ within the magnitude of the statistical noise.
Let $\underline{\mu }$ denote the value of the chemical potential when this
happens and save the state $S$ of simulation cell at that point.

\item  Then, gradually increase the chemical potential of the MC simulation
of phase $\alpha $ until its concentration can be predicted by the sequence $%
x_{i}^{\gamma }$ within the magnitude of the statistical noise. Let $%
\overline{\mu }$ denote the value of the chemical potential when this
happens.

\item  Restore the state of the simulation cell of phase $\alpha $ to the
saved state $S$ and set the chemical potential to $\mu =\left( \overline{\mu 
}+\underline{\mu }\right) /2$.

\item  Rerun the simulation for both phases $\alpha $ and $\gamma $ at
chemical potential $\mu $ and continue the integration Equation (\ref{dmudT}%
) with the values of $E$ and $x$ obtained for each phase.
\end{enumerate}

This approach results in a very robust algorithm that can run without user
intervention to map out the whole two-phase equilibrium. Thanks to this
safety mechanism, we found that it is feasible to integrate Equation (\ref
{dmudT}) towards increasing temperature, despite the potential for
instability with respect to noise. This proves useful since it is very easy
to find a starting point for our algorithm at $T=0$. The boundary tracing
algorithm may have to recenter itself a few times on its way to the end of
the two-phase equilibrium. But once the location of the phase boundary at
high temperature is known, a smooth phase boundary can then be calculated,
starting from the highest temperature and following the phase boundary in
the direction of decreasing temperature.

The end of the two-phase region can be detected by monitoring the occurrence
of either one the following events (see Figure \ref{termcrit}):

\illuseps{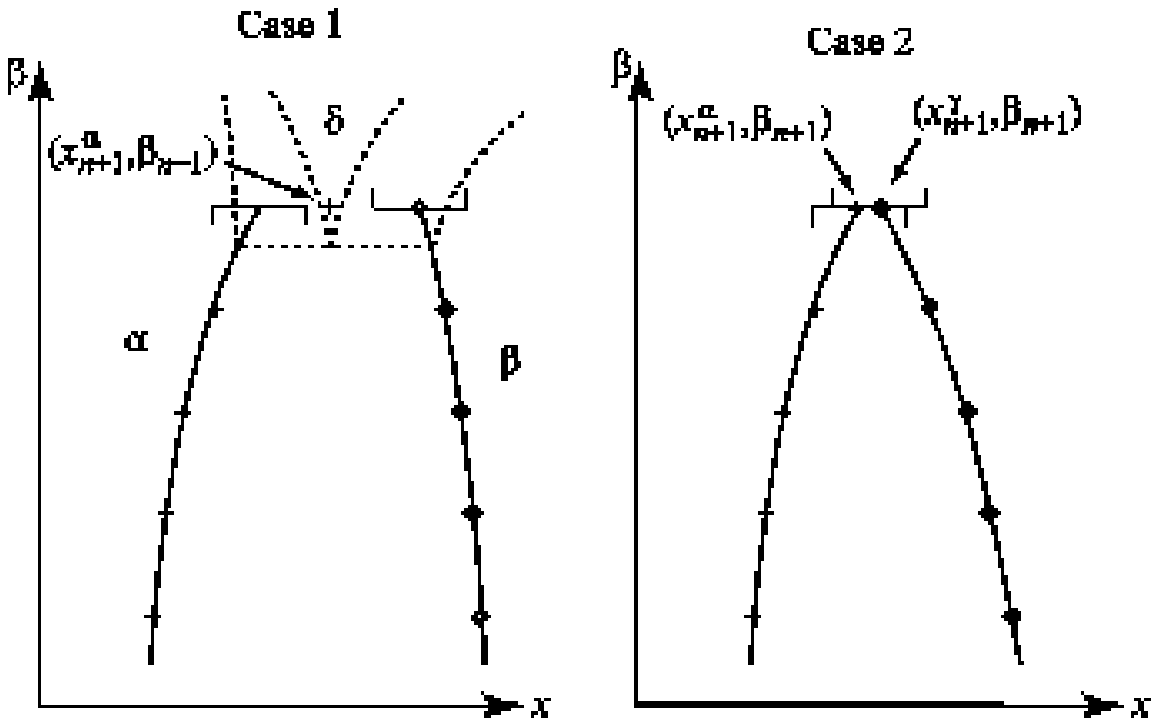}{Criterion for detecting the end of a
two-phase equilibrium. In Case 1, a new phase $\delta $ has appeared, as
indicated by the fact that $x_{n+1}^{\alpha }$ lies outside of the error bar
of the prediction of concentration for either phase $\alpha $ or $\gamma $.
In Case 2, a congruent or critical point has been reached since the
concentrations of the two phases are no longer statistically significantly
different.}

\begin{enumerate}
\item  The concentrations $x^{\alpha }$ and $x^{\gamma }$ are statistically
significantly different, but either $x^{\alpha }$ or $x^{\gamma }$ is not
predicted by the corresponding sequence $x_{i}^{\alpha }$ or $x_{i}^{\gamma
} $within statistical uncertainty.

\item  The concentrations $x^{\alpha }$ and $x^{\gamma }$ are not
statistically significantly different, but the sequences $x_{i}^{\alpha }$
and $x_{i}^{\gamma }$ both predict the same concentration, within
statistical uncertainty.
\end{enumerate}

Case 1 indicates that a third phase $\delta $ has become stable (i.e. a
eutectic or peritectic point), while Case 2 indicates that the phase
boundaries have met at a common composition (e.g. at a congruent point or a
critical point). In the first case, the algorithm will actually continue
slightly beyond the true eutectic or peritectic point\ because of the
presence of hysteresis at a first-order phase transition. However, the true
location of the three-phase equilibrium can be determined from the
intersection of the $\alpha -\delta $ and $\delta -\gamma $ phase
boundaries, once they have been determined.

\section{Applications}

This section presents a number of examples of automated calculations of
thermodynamic properties using the previously introduced algorithms. In each
example, the cluster expansion describing the energetics of the alloy system
was obtained from a fit to first-principles structural energy calculations
using the MAPS \cite{avdw:mapscode}\ code, which automates the process. A
description of the algorithms enabling the automatic construction of such
a cluster expansion, as well as a description of the interaction parameters
defining each cluster expansion used below, can be found in \cite{avdw:maps}.

Our first example is the determination of the complete potential surface $%
\phi \left( T,\mu \right) $ of the solid-state phases of the MgO-CaO
pseudo-binary alloy system (see Figure \ref{caomgop}). As the MgO-CaO
system is phase separating (as shown in Figure \ref{allphd}), the
thermodynamic potential surface exhibits a characteristic butterfly shape.
Below the critical temperature, the surface intersects with itself at a
nonzero angle, indicating the presence of a miscibility gap. Since the MC
simulation is able to sample metastable phases, for some values of the
chemical potential, one can determine $\phi \left( T,\mu \right) $ for both
phases in the region of metastability, which explains the double-valued
nature of the potential surface at low temperatures. Above the critical
temperature, $\phi \left( T,\mu \right) $ becomes single-valued and varies
smoothly as $\mu $ varies. (Note that, in reality, the alloy melts before
reaching that point at atmospheric pressure.) Figure \ref{caomgof} shows the
Gibbs\footnote{%
The difference between the Gibbs and Helmholtz free energies can be
neglected at atmospheric pressure in solid state systems because the product
of pressure $P$ with changes in specific volume ($\Delta V$) are small.}
free energy surface $G\left( x,T\right) \equiv \phi \left( T,\mu \left(
T,x\right) \right) +\mu \left( T,x\right) x$ for the same system and, in
this more familiar representation, the presence of miscibility gap at low
temperature is clearly recognizable. The calculated Gibbs free energy
surface can be directly used to fit standard solution models, such as an
ideal solution model, plus a polynomial in temperature and concentration to
account for non-ideality. The order of the polynomial needed to represent
the MC output can, for instance, be determined through cross-validation, as
defined in Section \ref{phtcv}.\ The resulting very compact representation
of the thermodynamics of the alloy system can be used in CALPHAD-type
calculations to supplement existing experimental data or to fill in data not
yet determined experimentally.

\illuseps{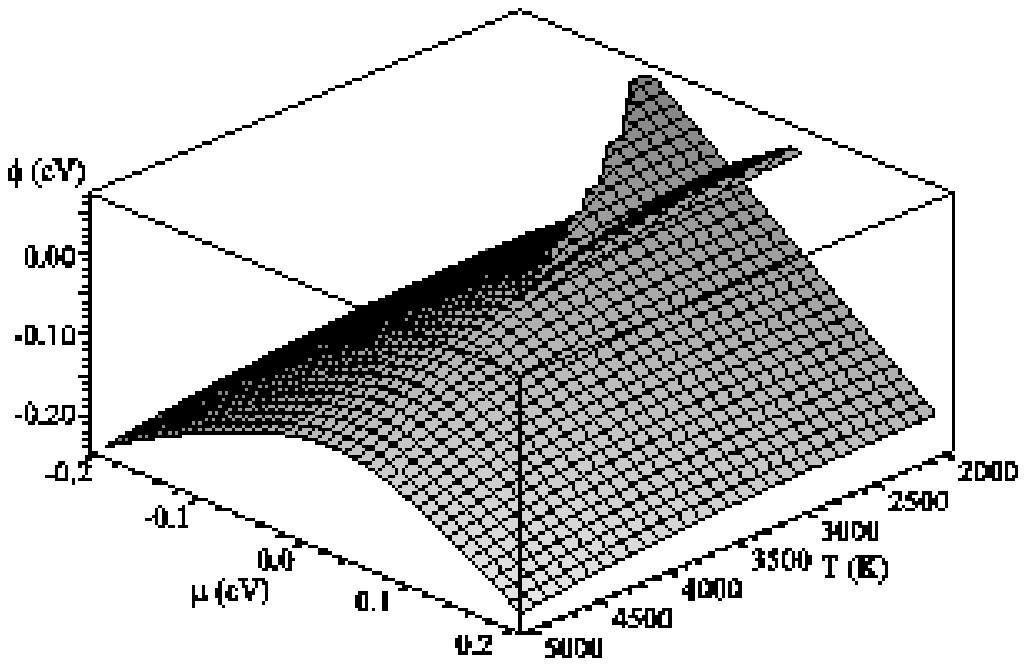}{Potential surface for the MgO-CaO
pseudobinary alloy system.}

\illuseps{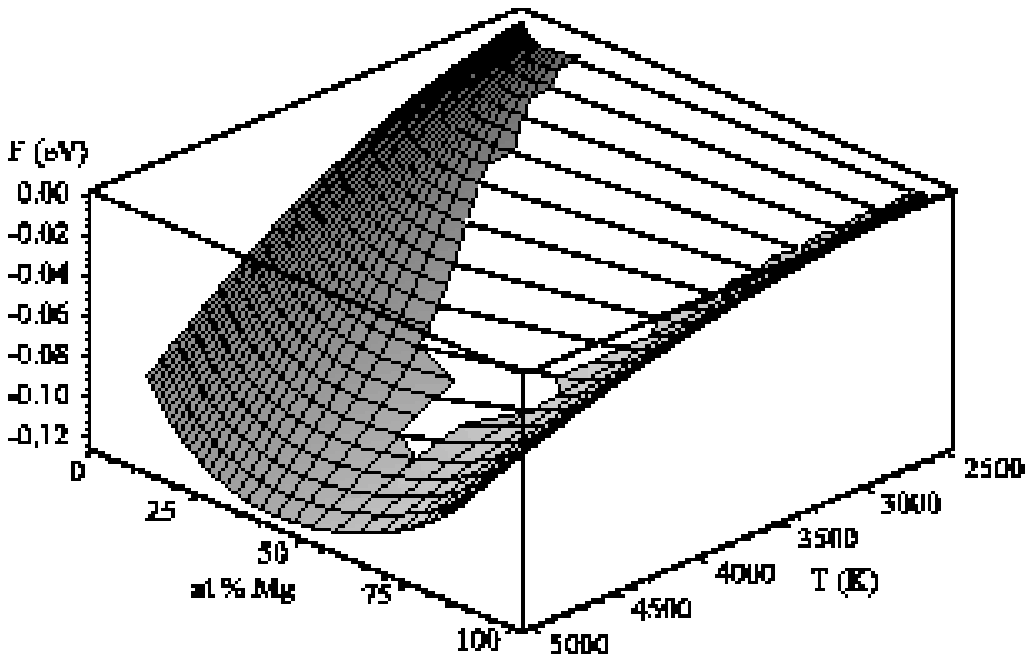}{Gibbs free energy for the MgO-CaO
pseudobinary alloy system. Common tangent constructions are shown at
selected temperatures.}

Figure \ref{tialp} shows the result of a similar calculation in the case
of a ordering alloy system, the Ti-Al system. For clarity, the metastable
portions of the $\phi \left( T,\mu \right) $ surface are omitted so that the
potential is single-valued everywhere. The order-disorder transition is
clearly visible from the kink in the surface, which is highlighted by a
thick curve. The corresponding Gibbs free energy representation of the same
data is shown in Figure \ref{tialf}.

\illuseps{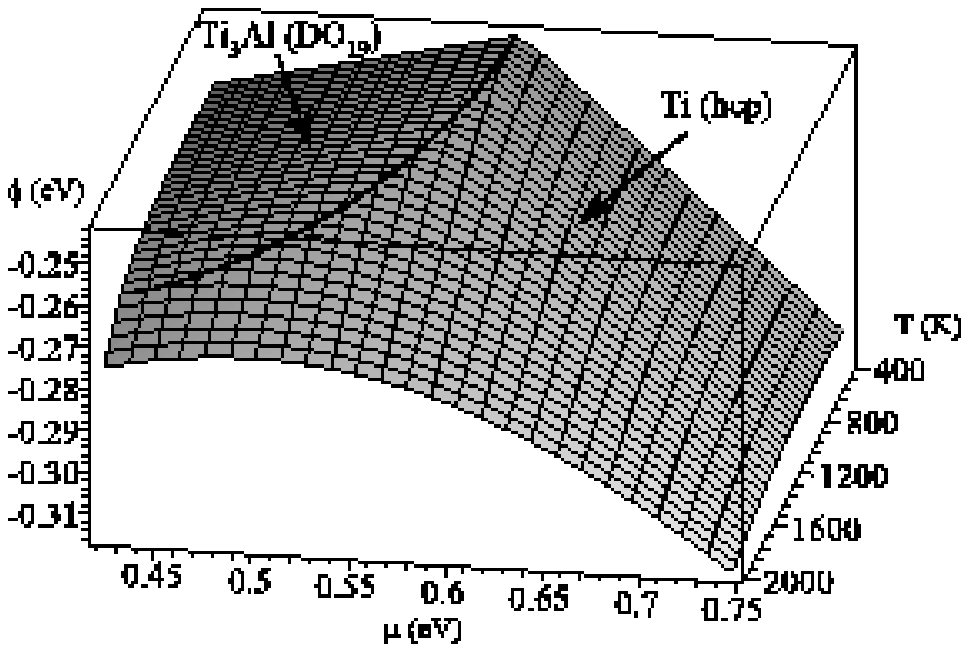}{Potential surface for the Ti-Al alloy system. The
thick curve indicates the location of a kink in the surface, associated with
a first-order transition.}

\illuseps{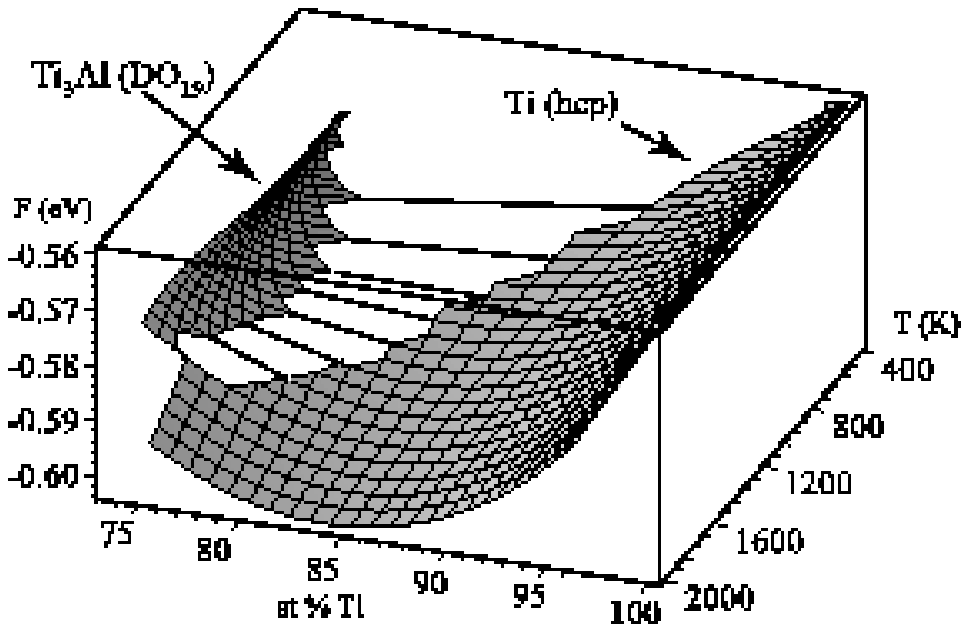}{Gibbs free energy for the Ti-Al alloy system.
Common tangent constructions are shown at selected temperatures.}

Figure \ref{allphd} illustrates the usefulness of the boundary tracing
algorithm described in Section \ref{phdtr}. The examples shown even include
a case (Ti-Al) where a equilibrium between two phases differing by their
lattice type: the L1$_{0}$ is fcc-based while the DO$_{19}$ and the Ti solid
solution are hcp-based). These examples show that one can obtain phase
diagrams from first-principles without first obtaining the complete $\phi
\left( T,\mu \right) $ surface.

\illuseps{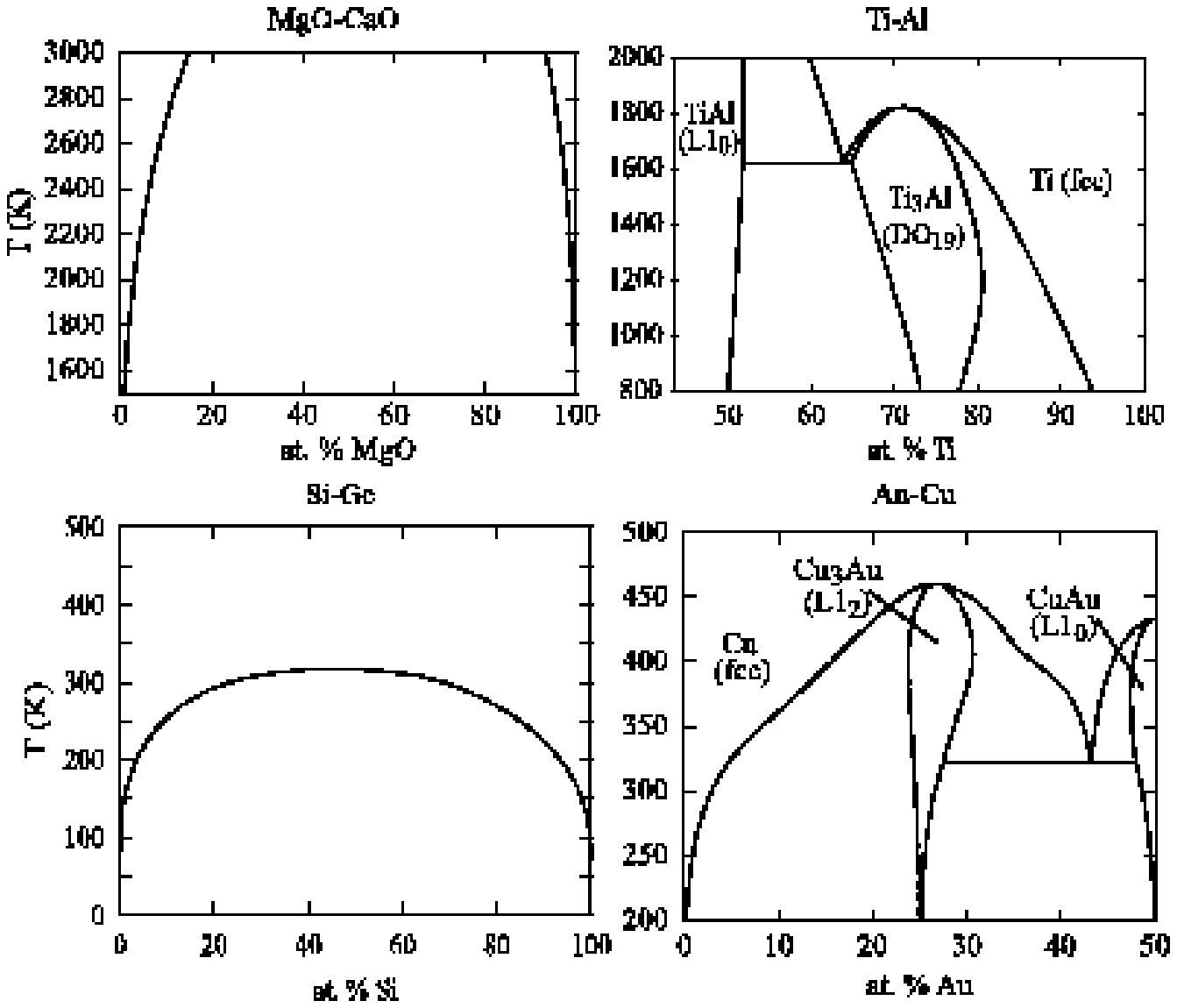}{Sample phase diagrams calculated with the
automatic boundary tracing algorithm.}

It is important to note that, in these examples, the MC code requires very
little intervention or input from the user. Thanks to the algorithms
introduced in Section \ref{autoprec}, the code is able to autonomously
determine, at each value of the control variables $\left( T,\mu \right) $,
when equilibrium has been reached and how long the simulation needs to be
run in order to reach the target precision. The code is also able to
determine when the region of stability of one phase has been exited, thanks
to the algorithms described in Section \ref{phtcv}, so that the user does
not need to specify when to stop the thermodynamic integration. The only
input parameters the user needs to specify are

\begin{enumerate}
\item  which phase to scan (the code can determine the value of the chemical
potential $\mu $ stabilizing the desired two-phase equilibrium at $0$K);

\item  the target precision (here, $0.1$ at \% precision on the
concentration variable is used as a criterion);

\item  the size of the simulation cell (in the present study, the simulation cell is
always chosen such that it contains a sphere of a diameter equal to at least 10 times the
nearest-neighbor interatomic spacing).
\end{enumerate}

\section{Conclusion}

The algorithms we have introduced enable researchers to focus on high-level
concepts when employing MC simulations. All the tasks that make traditional
Monte Carlo simulations so tedious have been formalized into algorithms that
can autonomously initialize and control the appropriate MC simulation.\ The
user only needs to specify high-level goals, such as the target precision or
which phase to focus on. This automation of MC simulations, along with the
earlier work that enabled the automation of cluster expansion construction 
\cite{avdw:maps}, signifies that it is now possible for large community of
researchers to employ first-principles calculations to augment and complete
existing experimental thermodynamic databases without facing the steep
learning curve that has traditionally been associated with first-principles
thermodynamical calculations.

\ack

This work was supported by the NSF under program DMR-0080766.


\section*{References}

\begin{thebibliography}{10}

\bibitem{newman:mc}
M.~E.~J. Newman and G.~T. Barkema, {\em Monte Carlo Methods in Statistical
  Physics} (Clarendon Press, Oxford, 1999).

\bibitem{binder:mc}
K. Binder and D.~W. Heermann, {\em Monte Carlo Simulation in Statistical
  Physics} (Springer-Verlag, New York, 1988).

\bibitem{dunweg:mc}
B. D{\"u}nweg and D.~P. Landau, Phys. Rev. B {\bf 48},  14182  (1993).

\bibitem{laradji:mc}
M. Laradji, D.~P. Landau, and B. D{\"u}nweg, Phys. Rev. B {\bf 51},  4894
  (1995).

\bibitem{sanchez:cexp}
J.~M. Sanchez, F. Ducastelle, and D. Gratias, Physica {\bf 128A},  334  (1984).

\bibitem{fontaine:clusapp}
D. de~Fontaine, Solid State Phys. {\bf 47},  33  (1994).

\bibitem{zunger:NATO}
A. Zunger,  in {\em NATO ASI on Statics and Dynamics of Alloy Phase
  Transformation}, edited by P.~E. Turchi and A. Gonis (Plenum Press, New York,
  1994), Vol.~319, p.\ 361.

\bibitem{jones:ldareview}
R.~O. Jones and O. Gunnarsson, Rev. Mod. Phys. {\bf 61},  689  (1989).

\bibitem{fontaine:soe}
D. de~Fontaine, Solid State Phys. {\bf 34},  74  (1979).

\bibitem{laks:recip}
D.~B. Laks, L.~G. Ferreira, S. Froyen, and A. Zunger, Physical Review B {\bf
  46},  12587  (1992).

\bibitem{ducastelle:book}
F. Ducastelle, {\em Order and Phase Stability in Alloys} (Elsevier Science, New
  York, 1991).

\bibitem{ceder:ising}
G. Ceder, Comput. Mater. Sci. {\bf 1},  144  (1993).

\bibitem{wolverton:releci}
C. Wolverton and A. Zunger, Phys. Rev. Lett. {\bf 75},  3162  (1995).

\bibitem{asta:noble}
M. Asta and S.~M. Foiles, Phys. Rev. B {\bf 53},  2389  (1996).

\bibitem{avdw:vibrev}
A. van~de Walle and G. Ceder, Rev. Mod. Phys. {\bf 74},    (2002), in press.

\bibitem{ozolins:alsc}
V. Ozoli{\c{n}}{\v{s}} and M. Asta, Phys. Rev. B {\bf 86},  448  (2001).

\bibitem{ozolins:cuau}
V. Ozoli{\c{n}}{\v{s}}, C. Wolverton, and A. Zunger, Phys. Rev. B {\bf 58},
  R5897  (1998).

\bibitem{wolverton:cual}
C. Wolverton and V. Ozoli{\c{n}}{\v{s}}, Phys. Rev. Lett. {\bf 86},  5518
  (2001).

\bibitem{wolverton:selec}
C. Wolverton and A. Zunger, Phys. Rev. B {\bf 52},  8813  (1995).

\bibitem{kikuchi:cvm}
R. Kikuchi, Phys. Rev. {\bf 81},  988  (1951).

\bibitem{lu:rsmc}
Z.~W. Lu, D.~B. Laks, S.-H. Wei, and A. Zunger, Phys. Rev. B {\bf 50},  6642
  (1994).

\bibitem{zunger:scord}
A. Zunger, MRS Bull. {\bf 22},  20  (1997).

\bibitem{wolverton:srorev}
C. Wolverton, V. Ozoli{\c{n}}{\v{s}}, and A. Zunger, J. Phys.: Condens. Matter
  {\bf 12},  2749  (2000).

\bibitem{ceder:oxides}
G. Ceder, A. van~der Ven, C. Marianetti, and D. Morgan, Modelling Simul. Mater
  Sci Eng. {\bf 8},  311  (2000).

\bibitem{asta:fppheq}
M. Asta, V. Ozolins, and C. Woodward, JOM - Journal of the Minerals Metals {\&}
  Materials Society {\bf 53},  16  (2001).

\bibitem{avdw:emc2code}
A. van~de Walle, {E}asy {M}onte {C}arlo {C}ode {(Emc2)}, 2001,
  http://cms.northwestern.edu/atat/.

\bibitem{afk:lte}
A.~F. Kohan, P.~D. Tepesch, G. Ceder, and C. Wolverton, Comp. Mat. Sci. {\bf
  9},  389  (1998).

\bibitem{woodward:sgce}
C. Woodward, M. Asta, G. Kresse, and J. Hafner, Phys. Rev. B {\bf 63},  094103
  (2001).

\bibitem{stone:cv}
M. Stone, J. Roy. Stat. Soc. B Met. {\bf 36},  111  (1974).

\bibitem{li:cv}
K.-C. Li, Ann. Stat. {\bf 15},  956  (1987).

\bibitem{goldberger:ols}
A.~S. Goldberger, {\em A Course in Econometrics} (Harvard University Press,
  Cambridge, Massachusetts, 1991).

\bibitem{avdw:mapscode}
A. van~de Walle, {MIT} {A}b-initio {P}hase {S}tability ({MAPS}) code, 2001,
  http://www.mit.edu/~avdw/maps/.

\bibitem{avdw:maps}
A. van~de Walle and G. Ceder, Automating First-Principles Phase Diagram
  Calculations, 2001, submitted to the Journal of Phase Equilibria.

\end{thebibliography}
\end{document}